\documentclass[11pt]{article}
\usepackage{latexsym}
\usepackage{amssymb}
\usepackage{amsmath}
\usepackage{graphicx}
\usepackage{colortbl}

\begin{document}

\def\Nset{\mathbb{N}}
\def\Ascr{\mathcal{A}}
\def\Bscr{\mathcal{B}}
\def\Cscr{\mathcal{C}}
\def\Dscr{\mathcal{D}}
\def\Escr{\mathcal{E}}
\def\Fscr{\mathcal{F}}
\def\Hscr{\mathcal{H}}
\def\Iscr{\mathcal{I}}
\def\Mscr{\mathcal{M}}
\def\Nscr{\mathcal{N}}
\def\Pscr{\mathcal{P}}
\def\Qscr{\mathcal{Q}}
\def\Rscr{\mathcal{R}}
\def\Sscr{\mathcal{S}}
\def\Wscr{\mathcal{W}}
\def\Xscr{\mathcal{X}}
\def\cupp{\stackrel{.}{\cup}}
\def\bold{\bf\boldmath}

\newcommand{\rouge}[1]{\textcolor{red}{ #1}}
\newcommand{\bleu}[1]{\textcolor{blue}{ #1}}
\newcommand{\boldheader}[1]{\smallskip\noindent{\bold #1:}\quad}
\newcommand{\PP}{\mbox{\slshape P}}
\newcommand{\NP}{\mbox{\slshape NP}}
\newcommand{\opt}{\mbox{\scriptsize\rm OPT}}
\newcommand{\ec}{\mbox{\scriptsize\rm OPT}_{\small\rm 2EC}}
\newcommand{\lp}{\mbox{\scriptsize\rm LP}}
\newcommand{\inn}{\mbox{\rm in}}
\newcommand{\deff}{\mbox{\rm sur}}
\newcommand{\MAXSNP}{\mbox{\slshape MAXSNP}}
\newtheorem{theorem}{Theorem}
\newtheorem{lemma}[theorem]{Lemma}
\newtheorem{corollary}[theorem]{Corollary}
\newtheorem{proposition}[theorem]{Proposition}
\newtheorem{definition}[theorem]{Definition}
\def\prove{\par \noindent \hbox{{\bf Proof}:}\quad}
\def\endproof{\eol \rightline{$\Box$} \par}
\renewcommand{\endproof}{\hspace*{\fill} {\boldmath $\Box$} \par \vskip0.5em}
\newcommand{\mathendproof}{\vskip-1.8em\hspace*{\fill} {\boldmath $\Box$} \par \vskip1.8em}

\definecolor{orange}{rgb}{1,0.9,0}
\definecolor{violet}{rgb}{0.8,0,1}
\definecolor{darkgreen}{rgb}{0,0.5,0}
\definecolor{grey}{rgb}{0.75,0.75,0.75}

\title {
\vspace*{-2cm}
{\huge Eight-Fifth Approximation for TSP Paths} \\[3mm]
}
\author{Andr\'as Seb\H{o}\footnote{CNRS, UJF, Grenoble-INP,  Laboratoire G-SCOP.
Supported by
the TEOMATRO grant ANR-10-BLAN 0207 ``New Trends in Matroids: Base Polytopes, Structure, Algorithms and Interactions''.}
}

\begingroup
\makeatletter
\let\@fnsymbol\@arabic
\maketitle
\endgroup

\begin{abstract}
We prove the approximation ratio $8/5$ for the metric $\{s,t\}$-path-TSP problem, and more generally for shortest connected $T$-joins.

The algorithm that achieves this ratio is the simple ``Best of Many'' version of Christofides' algorithm (1976), suggested by An, Kleinberg and Shmoys (2012), which consists in determining the best Christofides $\{s,t\}$-tour out of those constructed from  a family $\Fscr_{>0}$ of trees having a convex combination dominated by an optimal solution $x^*$ of the fractional relaxation.  They give the approximation guarantee $\frac{\sqrt{5}+1}{2}$ for such an $\{s,t\}$-tour, which is the first improvement after the   $5/3$ guarantee of Hoogeveen's Christofides type algorithm (1991).  Cheriyan, Friggstad and Gao (2012) extended this result to a $13/8$-approximation of shortest connected $T$-joins, for $|T|\ge 4$.

The  ratio $8/5$ is proved  by simplifying and improving the approach of An, Kleinberg and Shmoys that consists in completing $x^*/2$ in order to dominate  the cost of ``parity correction'' for spanning trees.  We partition the edge-set of each spanning tree in $\Fscr_{>0}$ into an $\{s,t\}$-path (or more generally, into a $T$-join) and its complement, which induces a decomposition of $x^*$. This decomposition can be refined and then efficiently used to complete  $x^*/2$ without using linear programming or particular properties of $T$, but by adding to each cut deficient for $x^*/2$ an individually tailored explicitly given vector, inherent in $x^*$.

A simple example shows that the Best of Many Christofides algorithm may not find a shorter $\{s,t\}$-tour than $3/2$ times the incidentally common optima of the problem and of its fractional relaxation.

\smallskip\noindent
\noindent{\footnotesize {\bf keywords}: traveling salesman problem, path TSP,
approximation algorithm, $T$-join, polyhedron}
\end{abstract}

\section{Introduction}\label{sec:i}

A Traveling Salesman wants to visit all vertices of a graph $G=(V,E)$, starting from his home  $s\in V$, and -- since it is Friday -- ending his tour at his week-end residence,  $t\in V$. 
Given the nonnegative valued length function $c:E\longrightarrow \mathbb{R}_+$, he is looking for a shortest $\{s,t\}$-tour, that is, one of smallest possible (total) length.

The Traveling Salesman Problem (TSP) is usually understood as the  $s=t$ particular case of the defined problem, where in addition every vertex is visited exactly once. This ``minimum length Hamiltonian circuit'' problem is one of the main exhibited problems of combinatorial optimization. Besides being \NP-hard even for very special graphs or lengths \cite{GarJT76}, even the best up to date methods of operations research, the most powerful computers programmed by the brightest hackers fail solving reasonable size problems exactly.

On the other hand, some implementations provide solutions only a few percent away from the optimum on some large ``real-life'' instances. A condition on the length function that certainly helps both in theory and practice is the triangle inequality. A nonnegative function on the edges that satisfies this inequality is called a {\em metric} function. The special case of the TSP where $G$ is a complete graph and $c$ is a metric is called the {\em metric TSP}. For a thoughtful and distracting account of the difficulties and successes of the TSP, see Bill Cook's book \cite{Coo12}.

If $c$ is not necessarily a metric function, the TSP is  hopeless in general: it is not only \NP-hard to solve but also to approximate, and even for quite particular lengths, since  the Hamiltonian cycle problem in $3$-regular graphs is \NP-hard \cite{GarJT76}. The practical context makes it also natural to suppose that $c$ is a metric.

\medskip
A {\em $\rho$-approximation algorithm} for a minimization problem is a polynomial-time algorithm that computes a solution
of value at most $\rho$ times the optimum, where $\rho\in\mathbb{R}$, $\rho\ge 1$.  The {\em guarantee} or {\em ratio} of the approximation is $\rho$.

\medskip
The first trace of allowing $s$ and $t$ be different is Hoogeveen's article \cite{Hoo91}, providing a Christofides type  $5/3$-approximation algorithm, again in the metric case. There had been no improvement until An, Kleinberg and Shmoys \cite{AKS12} improved this ratio to $\frac{1+\sqrt{5}}{2}<1.618034$ with a simple algorithm, an ingenious new framework for the analysis, but a technically involved realization.

The algorithm first determines an optimum $x^*$ of the fractional relaxation; writing $x^*$ as a convex combination of spanning trees and applying Christofides' heuristic for each, it outputs the best of the arising tours.  For the TSP problem $x^*/2$ dominates any possible parity correction, as Wolsey \cite{Wol80} observed, but  this is not true if $s\ne t$. However, \cite{AKS12} manages to perturb $x^*/2$, differently for each spanning tree of the constructed convex combination, with small average increase of the length.

 We adopt this algorithm and this global framework for the analysis, and develop new tools that  essentially change its realization and shortcut the most involved parts. This results in a simpler analysis guaranteeing a solution within $8/5$ times the optimum.

\medskip
We did not fix that the Traveling Salesman  visits each vertex exactly once, our problem statement requires only  that {\em every vertex is visited at least once.} This version has been introduced by Cornu\'ejols, Fonlupt and Naddef \cite{CFN85} and was called the {\em graphical TSP}. In other words, this version asks for the ``shortest spanning Eulerian subgraph'', and puts forward an associated polyhedron and its integrality properties, characterized in terms of excluded minors.

This version has many advantages: while the metric TSP is defined on the complete graph, the graphical problem can be sparse, since an edge which is not a shortest path between its endpoints can be deleted; however, it is equivalent to the metric TSP (see Tours below); the length function $c$ does not have to satisfy the triangle inequality; this version  has an unweighted special case, asking for the minimum size of a spanning Eulerian subgraph.

The term ``graphic'' or ``graph-TSP'' has eventually been taken by this all $1$ special case, that we do not investigate here and avoid these three terms used in a too diversified way, different from habits for other problems.  For comparison, let us only note the  guaranteed ratios for the cardinality versions of the problems: $3/2$ for the min cardinality of a spanning connected subgraph with two given odd degree vertices, and $7/5$ if all vertices are of even degree \cite{SV12}.


\section{Notation, Terminology and Preliminaries}

The set of non-negative real numbers is denoted by $\mathbb{R_+}$, $\mathbb{Q}$ denotes the set of rational numbers. We fix the notation $G=(V,E)$ for the input graph. For $X\subseteq V$ we write $\delta(X)$ for the set of edges with exactly one endpoint in $X$. If $w:E\longrightarrow \mathbb{R}$ and $A\subseteq E$, then we use the standard notation $w(A):=\sum_{e\in A} w(e)$.

\medskip\noindent
{\bf Tours}:  For a graph $G=(V,E)$  and $T\subseteq V$ with $|T|$ even, a {\em $T$-join} in $G$ is a set $F\subseteq E$
such that $T=\{v\in V: \hbox{$|\delta(v)\cap F|$ is odd}\}.$ For $(G,T)$, where $G$ is  connected, it is well-known and easy to see that a $T$-join exists if and only if $|T|$ is even \cite{LPL}, \cite{VYGENyellow}.
 A {\em $T$-tour} $(T\subseteq V)$ of $G=(V,E)$ is a set $F\subseteq 2E$ such that
\begin{itemize}
\item[(i)]$F$ is a $T$-join of $2G$,
\item[(ii)] $(V, F)$ is a connected multigraph,
\end{itemize}
where $2E$ is the multiset consisting of the edge-set $E$, and the multiplicity of each edge is $2$; we then denote  $2G:=(V,2E)$. It is not false to think about $2G$ as $G$ with a parallel copy added to each edge, but we find the multiset terminology better, since it allows for instance to keep the length function and its notation $c:E\longrightarrow \mathbb{R}_+$, or in the polyhedral descriptions to allow variables to take the value $2$ without increasing the number of variables; the length of a multi-subset will be the sum of the lengths of the edges multiplied by their multiplicities, with obvious, unchanged terms or notations: for instance the size of a multiset is the sum of its multiplicities; $\chi_A$ is the multiplicity vector of $A$;  $x(A)$ is the scalar product of $x$ with the multiplicity vector of $A$; a {\em subset of a multiset} $A$ is a multiset with multiplicities smaller than or equal to the corresponding multiplicities of $A$, etc. 

A {\em tour} is a $T$-tour with $T=\emptyset$.

When  $(G,T)$ or $(G,T,c)$ are given, we always assume without repeating, that $G$ is a connected graph, $|T|$ is even, and $c:E\longrightarrow \mathbb{R_+}$. The latter will be called the {\em length} function, $c(A)$ $(A\subseteq E)$ is the length of $A$. 
The {\em $T$-tour problem (TTP)} is to minimize the length of a $T$-tour  for $(G,T,c)$ as input. The subject of this work is the TTP for an arbitrary length function. 

If $F\subseteq E$, we  denote by $T_F$ the set of vertices incident to an odd number of edges in $F$; if $F$ is a spanning tree, $F(T)$ denotes the {\em unique $T$-join of $F$.}

The  {\em sum} of two (or more) multisets is a multiset whose multiplicities are the sums of the two corresponding multiplicities. If $X,Y\subseteq E$, $X+Y\subseteq 2E$ and $(V,X+Y)$ is a multigraph. Given $(G,T)$, $F\subseteq E$ such that $(V,F)$ is connected, and a $T_F\triangle T$-join $J_F$, the multiset $F + J_F$  is a $T$-tour; 
the notation ``$\triangle$'' stays for the {\em symmetric difference} (mod~$2$ sum of sets).

In \cite{SV12} $T$-tours were introduced under the term {\em connected $T$-joins}. (This first name may be confusing, since $T$-joins have only $0$ or $1$ multiplicities.)  Even if the main target remains $|T|\le 2$, the arguments concerning this case often lead out to problems with larger $T$.

By ``Euler's theorem'' a subgraph of $2G$ is a tour or $\{s,t\}$-tour  if and only if
its edges can be ordered to form a closed ``walk'' or a walk from $s$ to $t$,
that visits every vertex of $G$ at least once, and uses every edge as many times as its multiplicity.

\medskip
For  the TTP, a $2$-approximation algorithm is trivial by taking a minimum cost spanning tree $F$
and doubling the edges of a $T_F\triangle T$-join of $F$, that is, of $F(T_F\triangle T)$.

For $T=\emptyset$, Christofides \cite{Chr76} proposed  determining first a minimum length spanning tree $F$ to assure connectivity, and then to add to it a shortest $T_F$-join. The obvious approximation guarantee $3/2$ of this algorithm has not been improved ever since.  A {\em Christofides type algorithm} for general $T$ {\em adds a shortest $T_F\triangle T$-join instead.}

For $T=\{s,t\}$ $(s,t\in V)$ this has been proved to guarantee a ratio of $5/3$ by Hoogeveen \cite{Hoo91} and improved by An, Kleinberg and Shmoys \cite{AKS12}.  Hoogeveen's approach and ratio can be obviously extended to $T$-tours for arbitrary $T$ providing the same guarantee with a Christofides type algorithm and proof \cite[Introduction]{SV12} . In Section~\ref{sec:Results} we show an ``even more Christofides type'' proof, relevant for our improved ratio 8/5 (see Proposition). Cheriyan, Friggstad and Gao \cite{C12} provided the first ratio better than $5/3$ for arbitrary $T$, by extending the analysis of \cite{AKS12}, with extra work, different for $|T|\ge 4$, leading to the ratio $13/8=1.625$.

\medskip
{\em Minimizing the length of a tour or  $\{s,t\}$-tour is equivalent to the metric TSP problem or its path version} (with all degrees $2$ except $s$ and $t$ of degree $1$, that is, a shortest Hamiltonian circuit or path). Indeed, any length function of a connected graph can be replaced by a function on the complete graph with lengths equal to the lengths of shortest paths (metric completion): then a  tour or an $\{s,t\}$-tour can be ``shortcut'' to a sequence of edges with all inner degrees equal to $2$. Conversely, if in the metric completion we have a shortest Hamiltonian circuit or path we can replace the edges by paths and get a tour or $\{s,t\}$-tour.

Given $(G,T,c)$, the minimum length of a $T$-join in $G$ is denoted by $\tau(G,T,c)$. A {\em $T$-cut} is a cut $\delta(X)$ such that $|X\cap T|$ is odd. It is easy to see that a $T$-join and a $T$-cut meet in an odd number of edges. If in addition $c$ is integer, the maximum number of  $T$-cuts so that every edge is contained in at most $c$ of them is denoted by $\nu(G,T,c)$.  By a theorem of Edmonds and Johnson \cite{EdmJ73}, \cite{LPL} $\tau(G,T,c)=\nu(G,T,2c)/2$, and a minimum length $T$-join can be determined in polynomial time. These are useful for an intuition, even if we only use the weaker Theorem~\ref{thm:polyhedron} below. For an introduction and more about different aspects of $T$-joins, see
\cite{LPL}, \cite{SCHRIJVERyellow}, \cite{Fra11}, \cite{VYGENyellow}.

\medskip\noindent
{\bf Linear Relaxation}: We adopt the polyhedral background and notations of \cite{SV12}.

Let $G=(V,E)$ be a graph. For a partition $\Wscr$ of $V$ we introduce the notation
$$\delta(\Wscr) \ := \ \bigcup_{W\in\Wscr} \delta (W),$$
that is, $\delta(\Wscr)$ is the set of edges that have their two endpoints in different classes of $\Wscr$.

Let $G$ be a connected graph, and $T\subseteq V$ with $|T|$ even.
\begin{eqnarray*}
P(G,T) &\! := \!& \{x\in\mathbb{R}^{E}\!: \
 x(\delta(W)) \ge 2 \mbox{ for all } \emptyset\not=W\subset V \mbox{ with } |W\cap T| \hbox{ even,}\\
&& \hspace{2.3cm} x(\delta(\Wscr)) \ge |\Wscr| - 1 \mbox{ for all partitions $\Wscr$ of $V$,}\\
&&\hspace{2.3cm} 0\le x(e)\le 2 \hbox{ for all $e\in E$} \Bigr\} .
\end{eqnarray*}


Denote $\opt(G,T,c)$ the length of the shortest $T$-tour for input $(G,T,c)$. Let
$x^*\in P(G,T)$  minimize $c^\top x$ on $P(G,T)$.

\medskip\noindent
{\bf Fact}: Given $(G,T,c)$,   $\opt(G,T,c)\ge \min_{x\in P(G,T)}c^\top x=c^\top x^*$ .
\medskip

Indeed, if $F$ is a $T$-tour, $\chi_F$ satisfies the defining inequalities of $P(G,T)$.


The following theorem is essentially the same as Schrijver \cite[page 863, Corollary 50.8]{SCHRIJVERyellow}.

\begin{theorem}\label{thm:polytop}
Let $x\in\mathbb{Q}^E$ satisfy  the inequalities
\begin{eqnarray*}
& \hspace{0cm} x(\delta(\Wscr)) \ge |\Wscr| - 1 \mbox{ for all partitions $\Wscr$ of $V$,}\\
&\hspace{-1cm} 0\le x(e)\le 2 \hspace{0.2cm}\hbox{for all $e\in E$}.
\end{eqnarray*}
Then there exists a set $\Fscr_{>0}$,  $|\Fscr_{>0}|\le |E|$ of spanning trees and coefficients $\lambda_F\in\mathbb{R}, \lambda_F>0$, $(F\in\Fscr_{>0})$ so that
\[\sum_{F\in\Fscr_{>0}}\lambda_F=1, \qquad x\ge \sum_{F\in\Fscr_{>0}}\lambda_F\chi_F,\]
and for given $x$ as input, $\Fscr_{>0}$, $\lambda_F$ $(F\in\Fscr_{>0})$ can be computed in polynomial time.\end{theorem}

\prove  Let $x$ satisfy the given inequalities. If $(2\ge) x(e)>1$ $(e\in E)$, introduce an edge $e'$ parallel to $e$, and define $x'(e'):=x(e)-1$, $x'(e):=1$, and $x'(e):=x(e)$ if $x(e)\le 1$.  Note that the constraints are  satisfied for $x'$, and $x'\le \underline 1$. Apply Fulkerson's theorem \cite{F70} (see \cite[page 863, Corollary 50.8]{SCHRIJVERyellow}) on the blocking polyhedron of spanning trees: $x'$ is then a s convex combination of spanning trees, and by replacing $e'$ by $e$ in each spanning tree containing $e'$; applying then Carath\'eodory's theorem, we get the assertion. The statement on polynomial solvability follows from Edmonds' matroid partition theorem \cite{Edm70}, or the ellipsoid method \cite{GLS}.
\endproof

Note that the inequalities in Theorem~\ref{thm:polytop} form a subset of those that define $P(G,T)$.  In particular, any optimal solution $x^*\in P(G,T)$ for input $(G,T,c)$ satisfies the conditions of the theorem.  Fix  $\Fscr_{>0}$, $\lambda_F$ provided by the theorem for $x^*$, that is, \[\sum_{F\in\Fscr_{>0}}\lambda_F\chi_F\le x^*.\]
 {\em We fix the input $(G,T,c)$ and keep the definitions $x^*$, $\Fscr_{>0}$, $\lambda_F$ until the end of the paper.}

\medskip
It would be possible to keep the context of \cite{AKS12} for $s\ne t$ where metrics in complete graphs are kept and only Hamiltonian paths are considered (so the condition $x(\delta(v))=2$ if $v\ne s$, $v\ne t$ is added), or the corresponding generalization in \cite{C12} for $T\ne\emptyset$.  However, we find it  more comfortable to have in mind only $(G,T,c)$, where $c$ is the given function which is not necessarily a metric, and $G$ is the original graph that is not necessarily the complete graph, and without having a restriction on $T$. The paper can be read though with either definitions in mind, the only difference being  the use of $\sum_{F\in\Fscr_{>0}}\lambda_F\chi_F\le x^*$ without the irrelevant equality here to hold.

The reader can also substitute $T=\{s,t\}$ $(s,t\in V$ with $s=t$ allowed, meaning $T=\emptyset)$ for easier reading, none of the relevant features of the proofs will disappear.

\bigskip
Last, we state a well-known analogous theorem of Edmonds and Johnson for the blocking polyhedron of $T'$-joins in the form we will use it. (The notation $T$ is now fixed for our input $(G,T,c)$, and the theorem will be applied for several different $T'$ in the same graph.)

\begin{theorem}\label{thm:polyhedron} {\rm \cite{EdmJ73}, (cf. \cite{LPL}, \cite{SCHRIJVERyellow})}
Given $(G,T',c)$, ($T'\subseteq V$, $|T'|$ even, $c:E\longrightarrow\mathbb{R_+})$, let
\[Q_+(G,T') :=\{x\in\mathbb{R}^{E}\!: x(C) \ge 1 \mbox{ for each $T'$-cut $C$,}\,\, x(e)\ge 0 \hspace{0.2cm}\hbox{for all $e\in E$}\}.\]
  A shortest $T'$-join can be found in polynomial time, and if $x\in Q_+(G,T')$, $\tau(G,T',c)\le c^\top x$.
\end{theorem}

\medskip\noindent{\bf The guarantee of Christofides' algorithm for $T$-tours}

\smallskip
We finish the introduction to the $T$-tour problem with a proof of the $5/3$-approximation ratio for Christofides's algorithm. Watch the partition of the edges of a spanning tree into a $T$-join --if $T=\{s,t\}$, an $\{s,t\}$ path -- and the rest of the tree in this proof!  For $\{s,t\}$-paths this ratio was first proved by Hoogeveen \cite{Hoo91} slightly differently (see for $T$-tours in the Introduction of \cite{SV12}), and in \cite{fivethird} in a similar way,  as pointed out to me by David Shmoys.



\medskip\noindent
{\bf Proposition}: {\em Let $F$ be an arbitrary $c$-minimum spanning tree. Then
$\tau(G,T_F\triangle T,c)\le \frac{2}{3}\opt(G,T,c).$}

\smallskip
\prove $\{F(T), F\setminus F(T)\}$ is a partition of $F$ into a $T$-join and a $T\triangle T_F$-join (see Figure~\ref{fig:tree}). The shortest $T$-tour $K$ has a $T_F$-join $F'$ by connectivity, so $\{F', K\setminus F'\}$ is a partition of $K$ to a $T_F$-join and a $T_F\triangle T$-join.

If either $c(F\setminus F(T))\le \frac{2}{3}c(F)$ or $c(K\setminus F')\le \frac{2}{3}c(K),$  then we are done, since both
are $T\triangle T_F$-joins. If neither hold, then we use the $T\triangle T_F$-join $F(T)\triangle F'$. Since  $c(F(T))\le \frac{1}{3}c(F)\le \frac{1}{3}\opt(G,T,c)$ and $c(F')\le \frac{1}{3}c(K)=\frac{1}{3}\opt(G,T,c)$, we have $c(F(T)\triangle F')\le c(F(T)) + c(F')\le \frac{2}{3}\opt(G,T,c).$ \endproof

In the next section we exploit this simple argument in a more advanced context (see Proposition and its Corollary) that anticipates the proof of the main result.

\section{Results}\label{sec:Results}



\medskip
In this section we introduce the ``language'' of the paper, random sampling, that has been proved to be helpful for numerous problems. The ancestor of the method for the TSP can be viewed to be Wolsey's proof \cite{Wol80} of $\opt(G,\emptyset,c)\le 3/2c^\top x^*$, improved recently in the cardinality case by \cite{GhaSS11}, \cite{MomS11}, \cite{Muc12}, and for $T$-tours by \cite{AKS12}, \cite{C12}. Our use of probabilities here is only notational though, but an elegant notation does really help. In the second half of this section we state and prove the key lemmas.

The random sampling framework has been used by An, Kleinberg and Shmoys for TSP paths in a simple and original way with surprising success \cite{AKS12}. Readers familiar with \cite{AKS12} may  find helpful the explanations in Section~\ref{sec:connections} about the relation of the new results to this framework. In this section watch the new ideas contributed by the present work: the separation of $x^*$ into $p^*$ and $q^*$, and a further decomposition of $p^*$.


\noindent

\medskip
The coefficient $\lambda_F$ of each spanning tree $F\in\Fscr_{>0}$ in the convex combination dominated by  $x^*$ (see Theorem~\ref{thm:polytop}) will be {\em interpreted as a probability distribution of a random variable $\Fscr$, $$\Pr (\Fscr =F):= \lambda_F$$ whose values are spanning trees of $G$, and}
\[\Fscr_{>0}=\{ F\subseteq E : \hbox{$F$ spanning tree of $G$, }\Pr(\Fscr=F)>0\}.\]

The notations for spanning trees will also be used for random variables whose values are spanning trees. For instance $\Fscr (T)$ denotes the random variable whose value is $F(T)$ precisely when $\Fscr=F$. Another example is $\chi_\Fscr$, a random variable whose value is  $\chi_F$ when $\Fscr=F$. Similarly, $T_\Fscr$ is a random variable whose value for $\Fscr=F$ is $T_F:=\{v\in V: \hbox{$|\delta(v)\cap F|$ is odd}\}.$

\smallskip
We use now the probability notation for defining two vectors that will be extensively used:

\smallskip
$p^*(e):=\Pr(e \in \Fscr (T))$;\quad $q^*(e):=\Pr(e \in \Fscr\setminus \Fscr (T))$ $(e\in E).$
(These are short notations for the sum of $\lambda_F$ for spanning trees $F$ with $e \in F(T)$ or $e \in F\setminus F(T)$, respectively.)




\medskip\noindent
{\bf Fact}:  $E[\chi_{\Fscr(T)}]=p^*$, $E[\chi_{\Fscr\setminus \Fscr(T)}]=q^*$, $E[\chi_\Fscr]=p^*+q^*\le x^*$. \,\,\,{\bf Proof}: Apply Theorem~\ref{thm:polytop}. \endproof
\smallskip

Let us familiarize with the introduced vectors $p^*$, $q^*$ by sharpening the proposition at the end of the preceding section using the minimum objective  value of the fractional relaxation. This is irrelevant for the proofs in the sequel, but shows the  intuition of using $p^*$ and $q^*$.  

\medskip\noindent
{\bf Proposition}: {\em For each $T'\subseteq V$, $|T'|$ even, 
 $\frac{1}{2}(x^*+p^*)\in Q_+(G,T')$.}
\medskip

Let $\Qscr := \{\hbox {$Q$ is a cut: } x^*(Q) < 2 \}$. The assertion is that $p^*$ repairs the deficit of each $Q\in\Qscr$.

\medskip
\prove If $C$ is a cut, $C\notin\Qscr$, then  $x^*(C)\ge 2$, so $\frac{1}{2}(x^*(C) + p^*)\ge \frac{1}{2}x^*(C)\ge 1$. If $C\in\Qscr$:

$x^*(C)+p^*(C)\ge E[\chi_{\Fscr}](C)+ E[\chi_{\Fscr(T)}](C)\ge 2$, since the event $|C\cap\Fscr|=1$ implies that the unique edge of  $C\cap\Fscr$ is also contained in $\Fscr (T)$. (The $T$-cut $C$ intersects every $T$-join.)
 \endproof

 \begin{figure}[t]
 \vskip - 1.3cm
\includegraphics{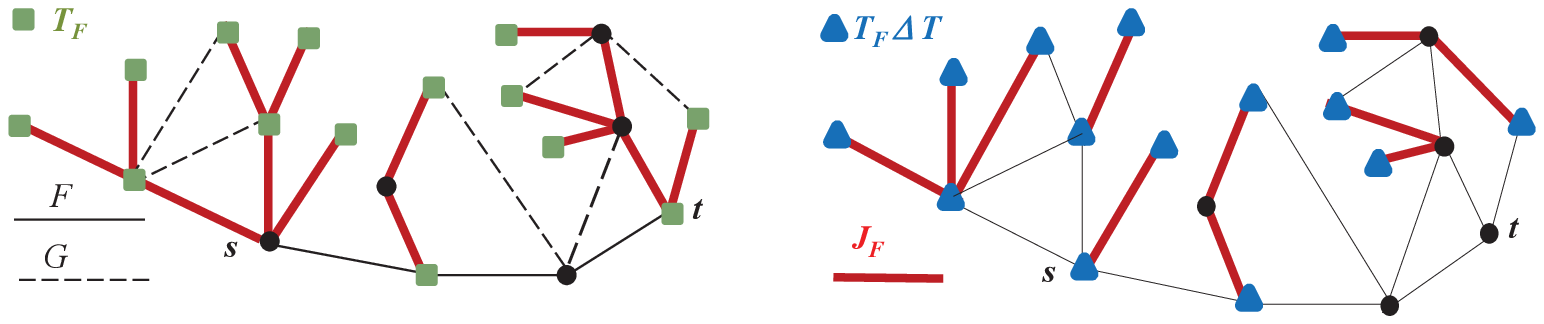}
\caption{\footnotesize One of many: \rouge{\bf $T_F\triangle T$-joins}, in $F$ (left), minimum in $G$ (right), \rouge{$J_F$};  $T:=\{s,t\}$. } \label{fig:tree}
\end{figure}

 \medskip\noindent
{\bf Corollary}: {\em
$\displaystyle E[\tau(G,T_\Fscr\triangle T,c)]\le \min\{c^\top x^*- \frac{c^\top q^*}{2}, c^\top q^*\}\le \frac{2}{3}c^\top x^*$.}   

 \prove Apply the Proposition and Theorem~\ref{thm:polyhedron} to get $\tau(G,T',c)\le c^\top \frac{1}{2}(x^*+p^*).$ Applying this to $T'=T_F\triangle T$ $(F\in\Fscr_{>0})$, and then substituting $\frac{1}{2}(x^*+p^*)\le x^* - \frac{1}{2}q^*$ (by the Fact), and finally taking the mean value: $E[\tau(G,T_\Fscr\triangle T,c)]\le c^\top x^*- \frac{c^\top q^*}{2}$.

 On the other hand, since $\Fscr \setminus \Fscr (T)$ is a $T_\Fscr\triangle T$-join,
  $E[\tau(G,T_\Fscr\triangle T,c)]\le E[c(\Fscr \setminus \Fscr (T))]= c^\top q^*$.

The minimum of our two linear bounds takes its maximum value at $c^\top q^*= \frac{2}{3} c^\top x^*$.
 \endproof
\medskip

While $c^\top  q^*$ (the second upper bound of the corollary) is the {\em mean value} of the length of the parity correcting $\Fscr \setminus \Fscr(T)$,   $\frac{1}{2}(x^*+p^*)$ (of the first bound) is in $Q_+(G,T')$ {\em for all} $T'\subseteq V$, $|T'|$ even. This ``for all'' is a superfluous luxury!  Indeed, it is not very economic to add $p^*$ for all $F\in\Fscr_{>0}$, when a smaller vector, adapted to $F$ (see below) is enough!

The reader may find helpful to have a look at Figure~\ref{fig:tree} for these remarks, for the following algorithm and for the subsequent arguments and theorem.

\medskip
\noindent
{\bf  Best of Many Christofides Algorithm \cite{AKS12}:}  Input $(G,T,c)$.

Determine $x^*$ \cite{GLS} using \cite{BC87}, see \cite{SV12}. (Recall: $x^*$ is an optimal solution of $\min_{x\in P(G,T)}c^\top x$.)

Determine $\Fscr_{>0}$.  (see Theorem~\ref{thm:polytop} and its proof.)

Determine the best {\em parity correction} for each $F\in \Fscr_{>0}$, i.e. a shortest  $T_F \triangle T$-join $J_F$  \cite{EdmJ73}, \cite{VYGENyellow}.

Output that  $F+J_F$ $(F\in \Fscr_{>0})$ for which $c(F + J_F)$ is minimum.

\bigskip
When $T=\emptyset$ $(s=t)$ Wolsey \cite{Wol80} observed that $x^*/2\in Q_+(G,T)$ and then by Theorem~\ref{thm:polyhedron} parity correction costs at most $c^\top x^*/2$, so Christofides's tour is at most $3/2$ times $c^\top x^*$; in \cite{AKS12}, \cite{C12} and here this  analysis is refined for paths and in general for $T$-tours.

\medskip
Define $\displaystyle R:= \min_{F\in \Fscr_{>0}} \frac {c(F)+ \tau(G,T_F\triangle T,c)}{c^\top x^*}\le \frac{E[c(\Fscr)+\tau(G,T_\Fscr\triangle T,c)]}{c^\top x^*}\le 1+E[\frac{\tau(G,T_\Fscr\triangle T,c)}{c^\top x^*}].$
Ratios of tour lengths versus $c^\top x^*$ may be better than $R$, since Christofides' way of choosing a spanning tree and adding parity correction is not the only way for constructing tours. For instance M\"omke and Svensson \cite{MomS11} get better results for some problems by starting from larger graphs than trees and deleting some edges instead  of adding them for parity correction. However, here we are starting with trees and correct their parity by adding edges for deducing the ratio $R\le 8/5$ through the following theorem, the main result of the paper:

\begin{theorem}\label{thm:main}
$\displaystyle E[\tau(G,T_\Fscr\triangle T,c)]\le \frac{3}{5}c^\top x^*$
\end{theorem}


Recall $\Qscr := \{\hbox {$Q$ is a cut: } x^*(Q) < 2 \}.$ {\em Every $Q\in \Qscr$ is a $T$-cut}, since non-$T$-cuts $C$ are required to have $x(C)\ge 2$ in the definition of $P(G,T)$.  In \cite{AKS12} it is proved that the vertex-sets defining $\Qscr$ form a chain if $|T|=2$; in \cite{C12} they are proved to form a laminar family for general $T$. We do not use these properties, but we need the following simple but crucial observation from \cite{AKS12}:

\begin{lemma}\label{lem:probound}
If $C$ is a cut, then $\Pr (|C\cap \Fscr | \ge 2)\le x^*(C) - 1,$ $\Pr (|C\cap \Fscr | =1)\ge 2 - x^*(C)$. Moreover if $C\in\Qscr$, then the event $|C\cap \Fscr | =1$ implies that $C$ is not a $T_\Fscr \triangle T$-cut.
\end{lemma}

\prove  If $C$ is a  cut of $G$, $x^*(C) \ge E[|C \cap \Fscr|] \ge \Pr (|C\cap \Fscr | = 1) + 2 \Pr (|C\cap \Fscr | \ge 2),$ where
$\Pr (|C\cap \Fscr | = 1) + \Pr (|C\cap \Fscr | \ge 2)=1,$ so the inequalities follow for an arbitrary cut. The last statement also follows, since  $C\in\Qscr$ implies that $C$ is a $T$-cut, and on the event $|C\cap \Fscr | =1$ it is also a $T_\Fscr$-cut --by degree counting--, so it is not a $T_\Fscr \triangle T$-cut, as claimed.
\endproof


It is for cuts $C\in\Qscr$ that this lemma provides relevant information. An, Kleinberg and Shmoys \cite{AKS12} need and prove more about $\Qscr$, their main technical tool \cite[Lemma~3]{AKS12} is actually a  linear programming fact about this family which is the more difficult half of their proof. Cheriyan, Friggstad, Gao \cite{C12} generalize these properties. The following two lemmas provide a natural simple alternative to this approach, inherent in the problem:

\begin{lemma}\label{lem:disjoint}
If $C_1\ne C_2$ are cuts of $G$, $e\in E$, then the events $\{e\}=C_1\cap \Fscr$ and $\{e\}=C_2\cap \Fscr$ are disjoint, and if they are $T$-cuts, these events are included in
the event $e\in \Fscr (T)$.
\end{lemma}
The statement is true for arbitrary cuts $C_1, C_2$, but it will be applied only for $C_1, C_2\in\Qscr.$

\medskip
\prove Indeed,  $\{e\}=C_1\cap F$ for some $F\in\Fscr_{>0}$ means that $e$ is the unique edge of $F$ in $C_1$, so  $C_1$ is the set of edges of $G$ joining the two components of $F\setminus \{e\}$. If $C_1\ne C_2$, then the event that  $\Fscr\setminus \{e\}$ defines $C_1$ or that it defines $C_2$, mutually exclude one another.

Moreover, if say $C_1$ is a $T$-cut, then it has a common edge with every $T$-join, so in the event $\{e\}=C_1\cap \Fscr$ we have  $e\in \Fscr(T)$, proving the last statement.
\endproof

For all $Q\in\Qscr$ and $e\in E$ define $x^Q(e):=\Pr (\{e\}=Q\cap \Fscr)$. In linear terms $x^Q\in\mathbb{R}^E$ is equivalently defined as
$$x^Q:=\sum_{F\in \Fscr_{>0}, |Q\cap F|=1} \lambda_F\chi_{Q\cap F}.$$

\begin{lemma}\label{lem:xC}
   Outside $Q$, $x^Q$ is $0$. Moreover,  $1^\top x^Q=x^Q(Q)\ge 2 - x^*(Q)$, and
 $$\sum_{Q\in\Qscr} x^Q \le p^*.$$
 \end{lemma}

 \prove  If $e\notin Q$, then $e\notin Q\cap F$ for all $F\in \Fscr_{>0}$, so $x^Q(e):=\Pr (\{e\}=Q\cap \Fscr)=0.$
 Now
 $$1^\top x^Q:=\sum_{F\in \Fscr_{>0}, |Q\cap F|=1} \lambda_F1^\top\chi_{Q\cap F}=\sum_{F\in \Fscr_{>0}, |Q\cap F|=1} \lambda_F1=\Pr (|Q\cap \Fscr | =1),$$
 so the first inequality  follows now from Lemma~\ref{lem:probound}. To see the second  inequality note that  for each $e\in E,$
 $$\sum_{Q\in\Qscr} x^Q(e) = \sum_{Q\in\Qscr}\quad\sum_{F\in \Fscr_{>0}, Q\cap F=\{e\}} \lambda_F = \sum_{Q\in\Qscr}\Pr (Q\cap\Fscr=\{e\}),$$
 and by Lemma~\ref{lem:disjoint} this is at most  $\Pr(e\in \Fscr (T))=p^*(e).$
  \endproof




\section{Proof}\label{sec:proof}

In this section we prove the promised approximation ratio (Theorem~\ref{thm:main}).  As  \cite{AKS12}, we want to complete the random variable  $\beta x^* + (1-2\beta)\chi_\Fscr,\,$ $1/3< \beta < 1/2$ to one that is in $Q_+(G,T_\Fscr\triangle T)$, by {\em adding a random variable}. The length expectation of what we get then is an upper bound for the price $\tau(G,T_\Fscr\triangle T,c)$ of parity correction, by Theorem~\ref{thm:polyhedron}.  The difficulty is to estimate the length expectation of the added random variable in terms of $c^\top x^*$.

Why just the form $\beta x^* + (1-2\beta)\chi_\Fscr\,$?  We follow \cite{AKS12} here: for all cuts $C\notin\Qscr$, that is, if $x^*(C)\ge 2$, we have then $\beta x^*(C) + (1-2\beta)\chi_\Fscr (C)\ge 2\beta + 1 - 2\beta =1$. By this choice it is sufficient to add  correcting vectors to $T_\Fscr\triangle T$-cuts in $\Qscr$, and we do not know of any alternative for this.

Why just in the interval $1/3< \beta< 1/2$ ?  We need $1-2\beta\ge 0$; $\beta\le 1/3$ 
would make the approximation ratio at least $5/3$.

\medskip
For any cut  $C$ we call  the random variable  $\max \{0, 1-(\beta x^*(C) + (1 - 2\beta)|C\cap\Fscr|)\}$ the {\em deficit} of $C$ for $\beta$, unless $C\in \Qscr$, $|C\cap\Fscr|=1$, when we define the deficit to be $0$ (see Lemma~\ref{lem:probound}).

\begin{lemma}\label{lem:def}  The deficit of a $T_\Fscr\triangle T$-cut $C$ for $\beta$ $(\beta\in(1/3,1/2))$ is constantly $0$, unless $C\in\Qscr$ and $|C\cap \Fscr|\ge 2$, and when it is positive, it is never larger than \[4\beta - 1 - \beta x^*(C).\]
\end{lemma}

Note that this value can be negative, but then the deficit of $C$ is constantly $0$. 


 \medskip
\prove
If $C\notin\Qscr$, then $x^*(C)\ge 2$, and we saw three paragraphs above that the deficit of $C$ for $\beta$ is $0$. If $C\in\Qscr$ then $C$ is a $T$-cut; if in addition $|C\cap \Fscr|=1$, then $C$ is also a $T_\Fscr$-cut, so it is not a $T_\Fscr\triangle T$-cut (Lemma~\ref{lem:probound}), and the deficit is defined to be $0$.

We proved: if $C$ is a  $T_\Fscr\triangle T$-cut and the deficit of $C$ for $\beta$ is not $0$, then $|C\cap \Fscr|\ge 2$. Substituting this inequality to the deficit: $1-(\beta x^*(C) + (1 - 2\beta)|C\cap\Fscr|)\le 4\beta -1 - \beta x^*(C).$
\endproof


Let $ f^Q(\beta):=\max \left\{0,\frac{ 4\beta - 1 - \beta x^*(Q)}{2- x^*(Q)} \right\}$, and $s^F(\beta):=\sum_{Q\in\Qscr, |Q\cap F|\ge 2} f^Q(\beta)x^Q$.


 \begin{lemma}\label{lem:sure} $\beta x^* + (1-2\beta)\chi_\Fscr + s^\Fscr (\beta) \in Q_+(G,T_\Fscr\triangle T)$ is the sure event for all $\beta\in(1/3,1/2).$
 \end{lemma}

 \prove By Lemma~\ref{lem:xC},  $x^Q(Q)\ge 2 - x^*(Q)$, so $f^Q(\beta)x^Q(Q)\ge 4\beta - 1 - \beta x^*(Q)$ by substituting the above definition of $f^Q(\beta)$.  On the other hand, by Lemma~\ref{lem:def}, the deficit of a $T_\Fscr\triangle T$-cut, if positive at all, is at most $4\beta - 1 - \beta x^*(Q).$
 \endproof

\noindent{\bf Theorem~\ref{thm:main}}\quad
$\displaystyle E [\tau(G,T_\Fscr\triangle T, c) ] \le \frac{3}{5} \, c^\top x^*.$

\begin{figure}[t]
\vskip -1.3cm
\includegraphics{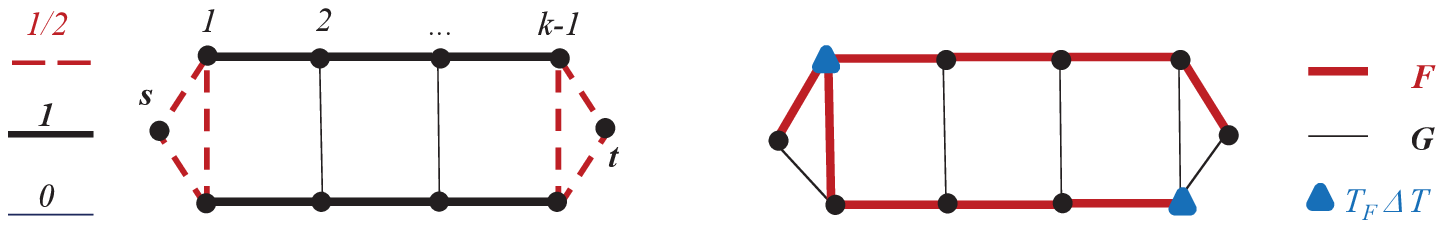}
\caption{\footnotesize The approximation guarantee cannot be improved below 3/2. This example is essentially the same as the more complicated one in \cite[ Figure~3]{SV12} providing the same lower bound for a more powerful algorithm in the cardinality case. $|V|=2k, \opt(G,T, \underline 1)=c^\top x^*=2k-1$ (left). Best of Many Christofides output (right): $3k-2$ if $\Fscr_{>0}$ consists of the thick (red) tree and its central symmetric image. There are more potential spanning trees for $\Fscr_{>0}$, but $\tau(G,T_F\triangle T,\underline 1)\ge k-2$ for each, so $c(F+J_F)\ge 3k-3$ for each, and with any $T_F\triangle T$-join $J_F$.}
\label{fig:threehalves}
\end{figure}

Figure~\ref{fig:threehalves} shows that this bound cannot be decreased below $1/2 \, c^\top x^*$.

\smallskip
\prove  Fix $\beta$, $1/3< \beta< 1/2$.

\smallskip\noindent{\bf Claim~1}:
$E[\tau(G,T_\Fscr\triangle T,c)]\le (1-\beta)c^\top x^* + c^\top E[s^\Fscr (\beta)]$ for all $1/3\le \beta\le 1/2$.

\medskip
By Lemma~\ref{lem:sure},
 $\beta x^* + (1-2\beta)\chi_\Fscr + s^\Fscr (\beta)\in Q_+(G,T_\Fscr\triangle T)$ is the sure event, so
$\tau(G,T_\Fscr\triangle T,c)\le c^\top(\beta x^* + (1-2\beta)\chi_\Fscr + s^\Fscr (\beta))$ also always holds. Taking the expectation of both sides and applying $E[c^\top(\beta x^* + (1-2\beta)\chi_\Fscr]\le(1-\beta)c^\top x^*$ (Fact of Section~\ref{sec:Results}), the Claim is proved.

\smallskip\noindent{\bf Claim~2}: For each $Q\in\Qscr$, $\displaystyle \Pr(|Q\cap \Fscr|\ge 2)f^Q(\beta)\le \frac{\beta\omega (3-\frac{1}{\beta} -\omega)}{1-\omega},$ where {\small $0\le \omega=1 - \sqrt{\frac{1}{\beta} - 2}<1$.}

By Lemma~\ref{lem:probound},  $\Pr(|Q\cap \Fscr|\ge 2)\displaystyle f^Q(\beta)\le (x^*(Q)-1)f^Q(\beta)\le\max_{Q\in\Qscr}(x^*(Q)-1)\frac{ 4\beta - 1 - \beta x^*(Q)}{2- x^*(Q)}.$
Substitute $\omega:=x^*(Q)-1$. Then the quantity to maximize becomes the function of $\omega$ in the claim. This function takes its maximum at the given value of $\omega$, and if $1/3\le\beta<1/2$ then $0\le \omega <1$, proving the Claim.

To be concise, denote $f(\beta):=\frac{\beta\omega (3-\frac{1}{\beta} -\omega)}{1-\omega},$ where $\omega=1 - \sqrt{\frac{1}{\beta} - 2}$.

\medskip\noindent{\bf Claim~3}:  $\displaystyle E[s^\Fscr (\beta)]\le f(\beta) p^*.$

\smallskip
$\displaystyle E[s^\Fscr (\beta)]= \sum_{F\in\Fscr}\Pr(\Fscr=F)\sum_{Q\in\Qscr, |Q\cap F|\ge 2} f^Q(\beta)x^Q=\sum_{Q\in\Qscr}\Pr(|Q\cap\Fscr|\ge 2)f^Q(\beta)x^Q\le$\newline
$\displaystyle\le  f(\beta) \sum_{Q\in\Qscr}x^Q,$ by Claim~2. Finally, substituting $\sum_{Q\in\Qscr} x^Q \le p^*$ (Lemma~\ref{lem:xC}) we get the claim.

\medskip
Now we are ready to finish the proof of the theorem. By Claim~1 and Claim~3, we have:
\[   E[\tau(G,T_\Fscr\triangle T,c)]\le (1-\beta)c^\top x^*  + f(\beta) c^\top p^*,         \]
where for all $\varepsilon\in\mathbb{R}, \varepsilon>0$, either $c^\top p^*\le (\frac{1}{2}-\varepsilon)c^\top x^*$, or $c^\top q^*\le (\frac{1}{2}+\varepsilon)c^\top x^*$ because $p^*+q^*\le x^*$ (Fact of Section~\ref{sec:Results}). So -- using the Fact again --, if the latter  case holds we have:
\[E[\tau(G,T_\Fscr\triangle T,c)]\le E[c(\Fscr\setminus \Fscr(T))]=c^\top q^*\le (\frac{1}{2}+\varepsilon)c^\top x^*,\]
and if the first case holds we can substitute $c^\top p^*\le (\frac{1}{2}-\varepsilon)c^\top x^*$ to the result we got before:
\[   E[\tau(G,T_\Fscr\triangle T,c)]\le (1-\beta)c^\top x^*  + (\frac{1}{2}-\varepsilon)f(\beta)c^\top x^*.\]
We got two upper bounds for $E[\tau(G,T_\Fscr\triangle T,c)]$, both having, for any fixed $\beta$, linear functions of $\varepsilon$ as coefficients of $c^\top x^*$. The minimum of the two functions has its maximum at $\varepsilon=\displaystyle \frac{1}{2} - \frac{\beta}{f(\beta)+1}$ which, as a function of $\beta$, has a unique minimum at  $\beta=4/9$ (and then $\omega=1/2$,  $f(\beta)=1/9)$, with minimum value $\varepsilon=1/10$.
\endproof


\section{Connections}\label{sec:connections}

Finally, we explain the connection of the results to their immediate predecessor, to some variants and to some open questions.

\noindent {\bf 5.1} First, we explain the content of this work in terms of An, Kleinberg and Shmoys \cite{AKS12}:

Replace the $\hat f_{U_i}^*$ provided by \cite[Lemma~3]{AKS12} -- whose existence is proved with linear-programming and network flow methods --  by the vector $x^Q$,  $x^Q(e):=\Pr (\{e\}=Q\cap \Fscr)$, see just above Lemma~\ref{lem:xC}. The
Lemma provides alternative simple properties for $x^Q$ that turn out to be more advantageous than those of $f_{U_i}^*$, moreover easy to prove.

The result of this change is that the maximum possible deficit $\beta\omega (\tau -\omega)$ of $T'$-cuts for a tentative `$T'$-join dominator', where $\omega=\tau/2$ (the place of the maximum) in \cite{AKS12},
is replaced by  $\displaystyle\frac{\beta\omega (\tau -\omega)}{1-\omega}$, where  $\omega=1 - \sqrt{\frac{1}{\beta} - 2}$ (the new place of the maximum), see Claim 2 of the proof.  

Another advantage is due to the fact that the new vectors sum up to a smaller vector than $c^\top x^*$: actually to at most $c^\top x^*/2$, and are  in fact dominated by $p^*$ (Lemma~\ref{lem:xC}), where $c^\top p^*<(\frac{1}{2}-\varepsilon)c^\top x^*$ unless $c^\top q^*<(\frac{1}{2}+\varepsilon)c^\top x^*$ (Fact in Section~\ref{sec:Results}).

\begin{figure}[!h]
\vskip -1cm
\includegraphics[bb=28 498 461 571]{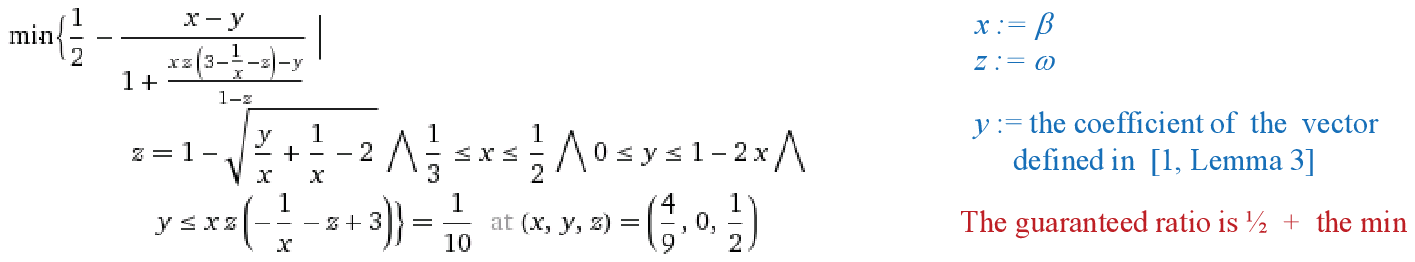}
\caption{\footnotesize  Mixing the performance of our analysis with that of An, Kleinberg, Shmoys \cite{AKS12}, optimally.} \label{fig:computation}
\end{figure}

 Despite these advantages, I cannot compare $\hat f_{U_i}^*$ and $x^Q$ directly. Therefore it seemed reasonable to hope that combining the two may further improve the bound $8/5$.  Figure~\ref{fig:computation} is the Wolfram Alpha output showing that this is not the case.

  If the coefficient of the sum of the $\hat f_{U_i}^*$ is $y$ -- this is the only single number that determines the extent of acting as \cite{AKS12} did --, our formulas in Section~\ref{sec:proof} are revised as follows.  In Lemma~\ref{lem:def} the upper bound becomes $4\beta - 1 - \beta x^*(Q) - y$ and then replacing Claim~1 and redoing Claim~3 accordingly (cf. the conclusion of these in the two lines following the proof of Claim~3), furthermore replacing $f^Q(\beta)$, $f(\beta)$ by the two-variable functions $f^Q(\beta,y)$, $f(\beta,y)$:
$$E[\tau(G,T_\Fscr\triangle T,c)]\le  (1-\beta +y)c^\top x^* + c^\top E[s^\Fscr (\beta)]\le (1-\beta +y)c^\top x^* + f(\beta,y)c^\top p^*,$$ with $\displaystyle \Pr(|Q\cap \Fscr|\ge 2)f^Q(\beta,y)\le \frac{\beta\omega (3-\frac{1}{\beta} -\omega) - y}{1-\omega}=:f(\beta,y)$, where $\omega=\sqrt{\frac{1}{\beta} - 2+\frac{y}{\beta}}$. We get now $\varepsilon=\displaystyle \frac{1}{2} - \frac{\beta - y}{f(\beta,y)+1}.$

 This is the function minimized in Figure~\ref{fig:computation}. According to Wolfram Alpha the minimum is reached for $y=0$, and then the result only confirms the hand computations of Section~\ref{sec:proof}.

 I was not able to exclude by hand that the minimum of this two-variable function  for $\varepsilon$  could be  smaller than $1/10$. Thanks to Louis Esperet and Nicolas Catusse for a pointer and a first guiding to Wolfram Alpha, that provided the answer of Figure~\ref{fig:computation}, and to Sebastian Pokutta who has double-checked the computations, with Mathematica. Of course, besides $\hat f_{U_i}^*$ and $x^Q$ there may be many other vectors to combine, and other possibilities for improvement.

\medskip\noindent
{\bf 5.2}
The results of the paper have obvious corollaries according to reductions of variants of the TSP to the TSP path problem, as a black box, for instance:

For the {\em clustered traveling salesman} problem \cite{fivethird} in which vertices of pairwise disjoint sets have to be visited consecutively, the update for the performance guarantee, where the number of clusters is a constant, is $8/5$; substituting our results  to \cite{AKS12}, we get that the {\em prize-collecting  $s$-$t$ path TSP problem,}  is $1.94837$-approximable.

\medskip\noindent
{\bf 5.3} Some of the questions that arise may be easier than the famous questions of the field:

Could the results of \cite{SV12} {\em $3/2$-approximating minimum size $T$-tours or $7/5$-approximating tours}  be reached {\em with the Best of Many Christofides} algorithm ?  Could the methods make the so far rigid bound of $3/2$ move down at least for {\em shortest $2$-edge-connected multigraphs} ?

\subsection*{Acknowledgment}
\small Many thanks to the organizers and participants of the  Carg\`ese Workshop of Combinatorial Optimization devoted to the TSP problem, for their time and interest, furthermore to Corinna, Jens, Kenjiro, Marcin  and Zoli Szigeti for their comments on this manuscript. I am highly indebted to Joseph Cheriyan, Zoli Kir\'aly and David Shmoys for their prompt and pertinent opinions   before my presentation, to R.~Ravi and Attila Bern\'ath, for their continuous interest and  wise suggestions.

\medskip
Thanks are also due to an anonymous pickpocket for a free day I could spend at Orly Airport, and to Easyjet for a delayed flight followed by a night I could spend at Saint Exup\'ery Airport. This research began, thanks to  their accidental, but helpful, day and night contributions.

\end{document}